\documentclass{iopart}
\usepackage{iopams}
\usepackage{graphicx}
\begin{document}
\title{Non-local coupling of two donor-bound electrons}

\author{J. Verduijn$^{1,2}$, R. R. Agundez$^2$, M. Blaauboer$^2$ and S. Rogge$^{1,2}$}
\address{$^1$ Centre for Quantum Computation and Communication Technology, University of New South Wales, Sydney NSW 2052, Australia}
\address{$^2$ Kavli Institute of Nanoscience, Delft University of Technology, Lorentzweg 1, 2628 CJ Delft, The Netherlands}
\ead{a.verduijn@unsw.edu.au}

\begin{abstract}
We report the results of an experiment investigating coherence and correlation effects in a system of coupled donors. Two donors are strongly coupled to two leads in a parallel configuration within a nano-wire field effect transistor. By applying a magnetic field we observe interference between two donor-induced Kondo channels, which depends on the Aharonov-Bohm phase picked up by electrons traversing the structure. This results in a non-monotonic conductance as a function of magnetic field and clearly demonstrates that donors can be coupled through a many-body state in a coherent manner. We present a model which shows good qualitative agreement with our data. The presented results add to the general understanding of interference effects in a donor-based correlated system which may allow to create artificial lattices that exhibit exotic many-body excitations.
\end{abstract}

\pacs{73.23.Hk,72.15.Qm,73.23.-b,85.35.Ds}

\maketitle

\section{Introduction}
Many physical effects in modern solid state physics are a manifestation of quantum interference. These effect, therefore, rely on the preservation of coherence of quantum states. Quantum dots embedded in an Aharonov-Bohm ring are ideal for studying coherent effects in solid state nano-structures \cite{Yacoby1995,Zaffalon:2008gb,Hatano:2011bn}. Shallow dopants in silicon have recently gained much interest because of their extremely long spin coherence times \cite{Steger:2012ev} and the reproducible confining potential \cite{Kane1998,Morton:2011vp}. For the same reasons that make them interesting for quantum device applications, single dopants are ideal to act as a flexible model system to investigate fundamental open problems in correlated systems in the solid state \cite{Lansbergen:2010bn,Tettamanzi:2012gr}. Here, we study a system of two dopants which are coherently coupled to contacts in a parallel configuration, thereby forming and an Aharonov-Bohm interferometer in a similar way to quantum dots embedded in a ring. We observe a peculiar type of Kondo effect mediating the interactions in the Coulomb-blockade regime of this system. And clarify this Kondo effect with a phenomenological model. \par
\begin{figure}[htbp]
\center
\includegraphics[width=90mm]{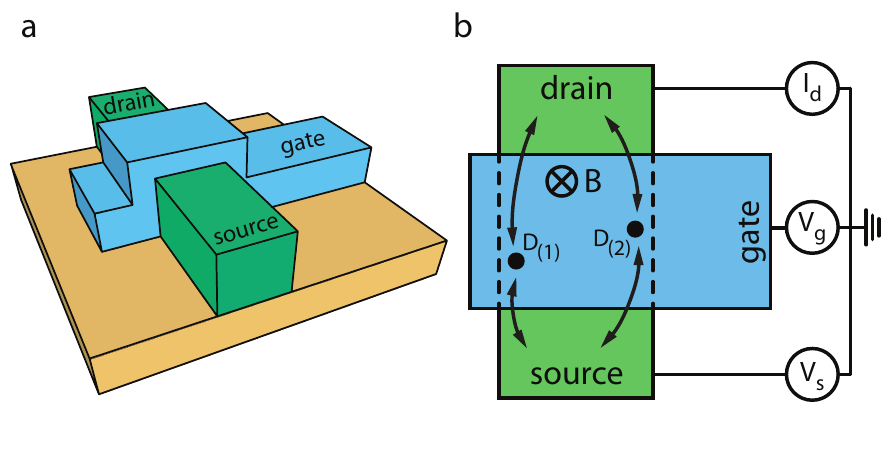}
\caption{a) To study single donor transport a three-dimensional FinFET device is used. Colored blue, the gate, wrapped around the channel (green), is shown. The device is fabricated on top of silicon dioxide (orange). b) In the channel region, below the gate, two arsenic donors in different corners of the channel (labeled D$_1$ and D$_2$) are coupled in parallel to the source and drain contacts. This arrangement is similar to an Aharonov-Bohm ring with a localized state in each arm. An applied magnetic field, $B$, induces a magnetic flux piercing through the loop enclosed by the conduction paths. Voltages are applied to source, $V_{s}$, and gate, $V_{g}$, while the drain current, $I_{d}$, is measured.}
\label{fig1}
\end{figure}

The system studied here consists of two arsenic donors in a field effect transistor that are coupled to leads in a parallel configuration, see Figure \ref{fig1}. This geometry allows for the tuning of the phase acquired by electrons as they traverse the structure. Changing the net magnetic flux threading the loop enclosed by the two conduction paths changes the acquired phase through the Aharonov-Bohm (AB) effect. In a previous publication we showed that this device exhibits phase-coherent transport in the sequential tunneling regime, evidenced by the presence of a Fano resonance \cite{Verduijn:2010ch}. Here, new data of the same device in the Kondo regime is presented and it is shown that the Kondo effect can be coherently modulated. \par
By studying the magnetic-field dependence of the Kondo transport, we link the phase modulation to the parallel arrangement of the donors. In a broader context, this system allows to study the rich behavior of occurrences of universal physical phenomena such as the Kondo effect \cite{GoldhaberGordon:1998vk}, the Aharonov-Bohm effect \cite{Aharonov1959} and the Fano effect \cite{Miroshnichenko:2010ew} in a mesoscopic system. \par
\section{Few-donor transport}
Single dopant transport spectroscopy has proven to be a powerful tool to study properties of dopants in nano-structures, e.g. \cite{Sellier2006,Tan:2010gi,Pierre:2010iu,Fuechsle:2012hx,Roche:2012gb}. Here, we use three-dimensional field effect transistors (FinFETs) with dopants embedded in the channel of the device as a platform for our experiments \cite{Sellier2006,Lansbergen2008,Pierre:2010iu}. Even though the channel is nominally only doped with boron, we occasionally find arsenic donors that have diffused into the p-type channel. We cannot control position of these dopants in the channel and, therefore, we rely on chance to find a dopant at a certain positions in the channel. The particular device we study here consists of two coupled dopants that dominate the sub-threshold transport, see Figure \ref{fig1}. All presented data has been obtained from a device which has a channel height of 60 nm, a gate length of 60 nm and channel width of 60 nm. Due to the strongly enhanced band-bending in the corner of the channel, close to the gate interface, the donors that show up in the sub-threshold transport are located in one of the corners of the channel \cite{Sellier:2007tb,Lansbergen2008}. The total cross-section of these corner channels is typically about 4 nm$^2$, as measured by thermally activated transport \cite{Sellier:2007tb}. Given the effective inter-donor distance of a few tens of nanometers, inferred from sequential tunneling data \cite{Verduijn:2010ch}, it is likely that the two donors in our device are located in different corners, allowing for a magnetic flux to pierce through the area enclosed by the conduction paths. This scenario is consistent with both the magnetic field dependence of the observed Fano resonance, and the Kondo effect, as we will show below. Thus, it is really due to the corner effect that we are able to perform transport spectroscopy on few-donor systems with a relatively large inter-donor separation at low temperature  ($\lesssim4$ K). \par
Transport spectroscopy on coupled dopants has been report in several systems. Evidence of dopant molecules was observed in the magnetic field dependent current in resonant tunneling diodes \cite{Geim:1994jy}. Embedded in a transistor, a small number of gated coupled donors were found to result in time-dependent fluctuations in the current which were attributed to multi-electron tunneling events \cite{Kuznetsov:1996fv}. Fano resonances, caused by closely spaced coupled dopants, were observed in a Schottky barrier metal oxide semiconductor FET \cite{Calvet:2011ht}. Direct resonant tunneling was recently observed in a pair of phosphorus donors \cite{Roche:2012gb}. In this work we demonstrate that, co-existing with interference in the sequential tunneling regime \cite{Verduijn:2010ch}, the current in the Coulomb blockade regime of a pair of coupled donors is showing interference due to many-body effects. \par
We measure the DC differential conductance $G=\textrm{d}I/\textrm{d}V_{{sd}}$ as a function of magnetic field, $B$, gate voltage $V_{g}$ and source/drain bias voltage, $V_{sd}$. Figure \ref{fig2}a shows a conductance trace close to zeros source/drain bias voltage. Two arrows indicate the position of the observed Fano resonances. Since the signal from the D$^0_{(1)}$ Fano resonance is very weak, we focus on the D$^-_{(1)}$. The hallmarks of single donor transport are a large charging energy and odd/even spin filling of the donors \cite{Sellier2006}; both have been observed in the device presented here. We perform all measurements at 0.3 K and $V_{sd} = 50\,\mu$V except when explicitly mentioned otherwise. In this article we focus on the physics governing the transport when both donors in the device are nominally occupied by one electron (near the circles in Figure \ref{fig2}b) and on the Fano resonance labeled D$^-_{(1)}$ in Figure \ref{fig2}. \par
\section{Aharanov-Bohm effect in the Kondo regime}
\begin{figure}[htbp]
\center
\includegraphics[width=120mm]{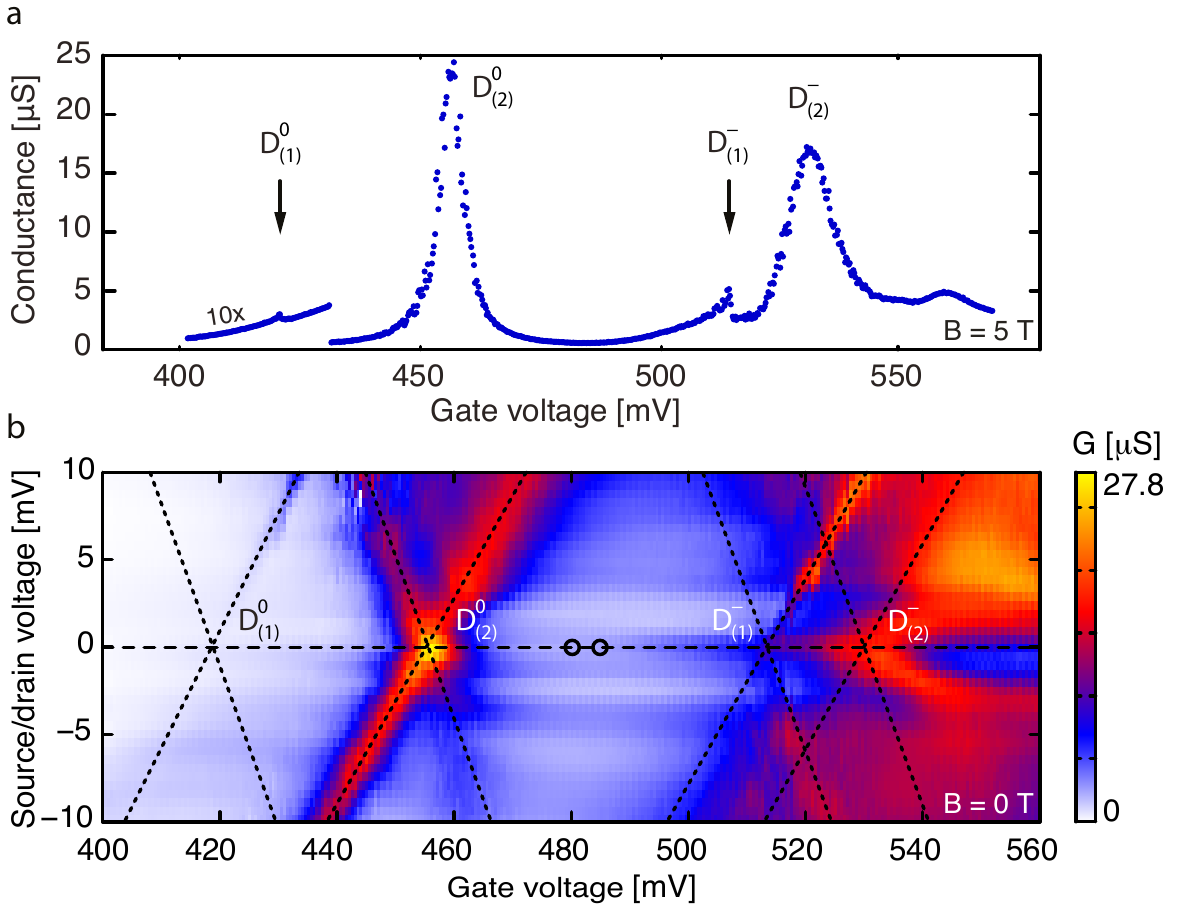}
\caption{a) The conductance at low source/drain voltage ($V_{sd} = 50\,\mu$V) shows four resonances due to the D$^0$ and D$^-$ states of two donors, (1) and (2), \cite{Lansbergen:2011eu}. Fano resonances result from the interference of the resonance channels induced by donor (1) and an approximatly continuum channel, labeled D$^0_{{(1)}}$ and D$^-_{{(1)}}$ (see main text). b) A differential conductance charge stability diagram shows that there is considerable conductance around zero source/drain voltage between the resonances labeled D$^0_{{(2)}}$ and D$^-_{{(1)}}$, which we attribute to a Kondo effect. As a guide to the eye the dotted lines that are labeled in the same way as in panel a) denote the resonances as finite bias. The black circles denote the gate voltages where the temperature dependence and magnetic field dependence of the transport have been investigated.}
\label{fig2}
\end{figure}

In the Coulomb-blockade region, where both donors are occupied by a single electron, we observe considerable zero-bias conductance, see Figure \ref{fig2}b. This is unlikely due to thermal effects, since the charging energy of the donors is large compared to the thermal energy $k_{B}T$ at the experimental temperature, i.e. $U/2k_{B}T\approx 500$. Zero-bias conductance is one of the characteristics of the spin-1/2 Kondo effect in single donors \cite{Lansbergen:2010bn} and quantum dots \cite{GoldhaberGordon:1998vk,Cronenwett:1998wc}. To see if this is indeed the cause of the conductance in the Coulomb blockade region, we investigate the temperature and magnetic field dependence of the conductance at a gate voltage of $V_{g}=485$ mV and $V_{g}=480$ mV respectively, see Figure \ref{fig2}a and b. For the Kondo effect in quantum dots the conductance increases logarithmically as the temperature is $\lesssim T_{K}$ and saturates at a zero-temperature maximum as the temperature $\ll T_{K}$, where $T_{K}$ is the Kondo temperature. Figure \ref{fig3}b shows a fit of a phenomenological scaling relation \cite{GoldhaberGordon:1998vk} to the temperature dependent data which describes this behavior. The Kondo temperature resulting from this fit is $T_{K}=(12\pm5)$ K, a value that justifies that we use an effective zero-temperature model to analyze the conductance later on. Figure \ref{fig3}c shows a plot of the measured differential conductance as a function of source/drain bias voltage and magnetic field. Since the transport processes in the Kondo regime involve spin-flip tunneling transitions, a gap $\sim2g\mu_{{B}}B$ wide (with $g=2$ for silicon), centered at zero bias, is expected to open \cite{GoldhaberGordon:1998vk,Meir1993}. This is indeed what we observe in the data, see Figure \ref{fig3}c. At a gate voltage of $V_{g}=490$ mV) we plot the Kondo conductance as a function of magnetic field and show that the Kondo effect (zero bias conductance) is quenched and a gap open as the magnetic field increases, see also the corresponding trace in Figure \ref{fig3}a. This has been recognized as one of the hallmarks of a Kondo effect in similar systems \cite{GoldhaberGordon:1998vk}. \par
There is, however, a discrepancy between our data and a simple single impurity spin-Kondo effect, which we believe is due to the magnetic field dependence of the phase, which is carried by electrons and tuned through the Aharonov-Bohm effect. Figure \ref{fig3}a shows that the zero-bias conductance unexpectedly increases with increasing field and subsequently decreases again, exhibiting a maximum at $B\sim7.7$ T. We find that this effect persists throughout the Coulomb-blockade region and the position of the maximum is almost independent of gate voltage. A possible explanation is that the Kondo tunneling processes are constructively interfering at this particular magnetic field and therefore cause an enhancement of the conductance. In the next section we investigate this idea further. \par
\section{Results and discussion}
In a Aharonov-Bohm interferometer the transport phase acquired by electrons traversing the structure contains a flux-dependent component in addition to a (constant) geometric phase \cite{Yacoby1995,Kobayashi:2002cq}. From transport measurements in the sequential tunneling regime we infer that the donors in this device are separated by a few tens of nanometers \cite{Verduijn:2010ch}, schematically depicted in Figure \ref{fig1}b. As mentioned before, this geometry is effectively a parallel configuration where a finite magnetic flux can pierce the area enclosed by the conduction paths and thus tunes the relative phase of the electrons by the Aharonov-Bohm effect \cite{Aharonov1959}. Depending on the total effective area enclosed by the conduction paths at a given magnetic field, electrons tunneling via one of the donors acquire a different phase compared to electrons tunneling via the other donor. By changing the magnetic field this phase can be tuned. As we will argue below, this magnetic field dependent phase affects not only electrons tunneling sequentially, which are effectively non-interacting, but also correlated electrons in the Coulomb blockade regime. \par
\subsection{Sequential transport in relation to the Kondo effect}
The effect of the magnetic field dependent phase due to the AB effect shows up in the a symmetry change of the D$^-_{(1)}$ resonance (Figure \ref{fig2}a) with a Fano line shape in gate voltage as a function of magnetic field. In the sequential transport regime, the interference of transport channels induced by the two donors, (1) and (2), causes this Fano interference effect \cite{Verduijn:2010ch}. The transport channels involved in this case are resonant sequential transport around the degeneracy point of the D$^0_{(1)}$ and D$^-_{(1)}$ (labeled D$^-_{(1)}$ in Figure \ref{fig2}) and a non-resonant channel (with a slow phase variation) generated by donor (2). 
We have determined, for the same device, that the period associated with this form of the Aharonov-Bohm effect is $B_{AB}\sim6.5$T \cite{Verduijn:2010ch}. We will show that this is consistent with the observed interference effect in the Kondo regime. \par
\begin{figure}[htbp]
\center
\includegraphics[width=120mm]{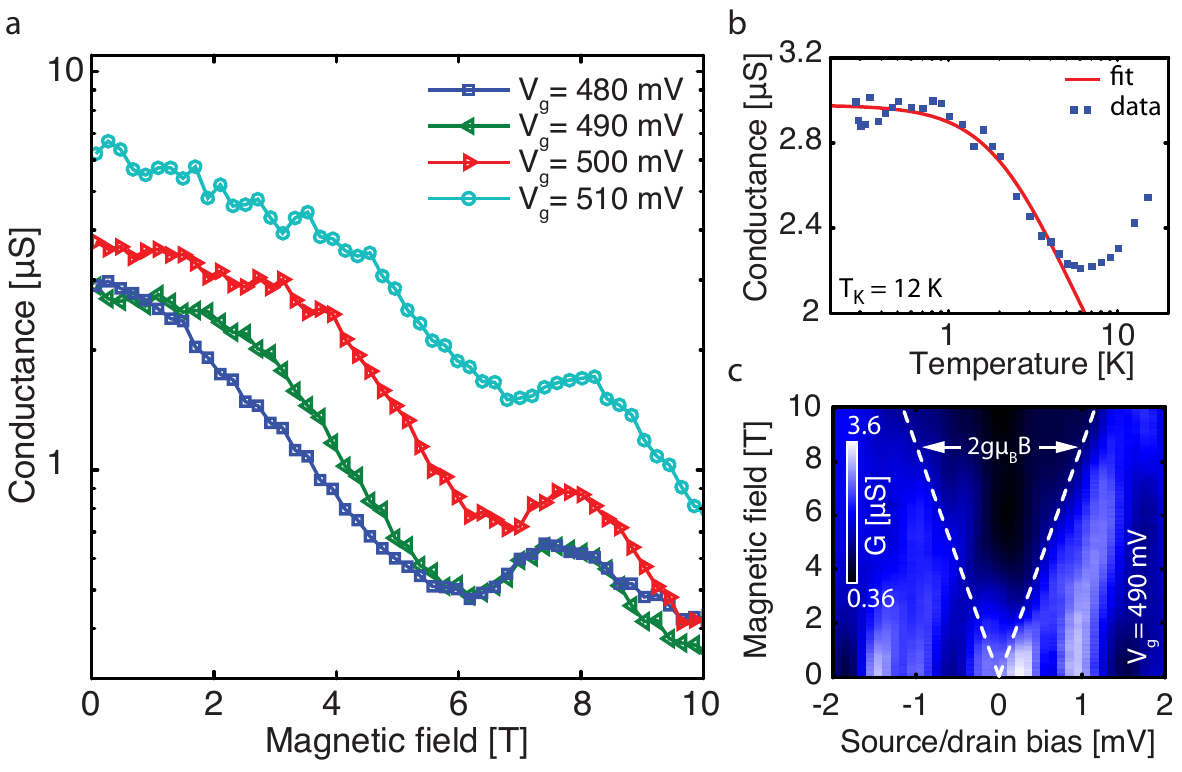}
\caption{The Kondo conductance is measured in the Coulomb-blockade region at several gate voltages as a function of magnetic field, temperature and source/drain bias voltage. a) At $\sim6$ T a clear upturn, with a maximum at $\sim7.7$ T, of the conductance as the magnetic field increases is visible. This is attributed to constructive interference of Kondo channels induced through both dopants, see also the discussion in Section 4. b) The temperature dependence of the Kondo conductance at $V_{g}=485$ mV. A fit of a semi-empirical spin-1/2 Kondo model for $T<5$ K has been performed, with the scaling parameter left free (red line) \cite{GoldhaberGordon:1998vk}. This fit results in a Kondo temperature of 12 K and scaling parameter $0.23\pm0.1$, consistent with a spin-1/2 Kondo effect. c) The Kondo conductance as a function of magnetic field and source/drain bias voltage at $V_{g}=490$ mV. A zero-bias gap, characteristic for a Kondo effect, opens when a magnetic field is applied.}
\label{fig3}
\end{figure}
Now, we turn back to the data in the Coulomb-blockade region shown in Figure \ref{fig3}a. These data show a change in the conductance as a function of magnetic field, very similar to conventional AB oscillations in mesoscopic systems \cite{Yacoby1995,VanderWiel:2000uw}, the main difference being that the magnitude decreases with magnetic field. In the conventional AB effect, as observed in mesoscopic rings, the interference alternates between constructive and destructive as the field is swept, depending on the magnetic field dependent phase picked up as electrons traverse the circumference of a loop. In our experiment, however, the continuum conduction paths are formed by spin-Kondo channels in both arms. This would explain why the amplitude of the oscillation decreases with magnetic field since the spin Kondo transport is quenched when a magnetic field is applied \cite{GoldhaberGordon:1998vd}. Note that a transport channel, connecting source and drain, and interfering with the donor-induced channel is not consistent with the observed gate-voltage current characteristics. Such a parallel channel would result in a monotonous background that increases with gate voltage, which we do not observe in the data in Figure \ref{fig2}. Instead, the sub-threshold conductance in dominated by sequential tunneling resonances with Coulomb blockade valleys between them. Another mechanism, which could potentially enhance the conductance in the Kondo regime at finite field, is the crossing of spin singlet and triplet states at finite magnetic field \cite{VanderWiel:2000uw}. This possibility can also be ruled out since this would also be visible as a kink in the shift of the Coulomb-blockade peaks \cite{Weis:1993ey} which we do not observe for the magnetic field range covered in our experiment, $B<10$ T. To further quantify the proposed double-donor induced Kondo effect, we develop a model in the next section. \par
\subsection{Interfering Kondo channels}
Taking a phenomenological approach, we derive an analytical expression for the Kondo conductance using a slave-boson mean-field approximation within a scattering matrix formalism. Assuming that both donors are fully in the Kondo regime and neglecting inter-donor Coulomb interaction, an expression for the magnetic-field dependent conductance can be derived. The assumption of vanishing inter-donor Coulomb interaction is supported by the fact that at large source/drain bias voltage (not shown here), where the donor resonances cross, we do not observe a significant shift of the respective resonances \cite{Pierre:2010iu}. Therefore, the zero-field Kondo effect can be modeled, for each donor independently, as a single resonance at the Fermi energy at zero temperature, i.e. $T\ll T_{K}$, as can be seen in Figure \ref{fig2}b. The latter assumption is confirmed by slave-boson mean-field calculations we have performed for parameters similar to the experimental conditions. \par
\begin{figure}[htbp]
\center
\includegraphics[width=120mm]{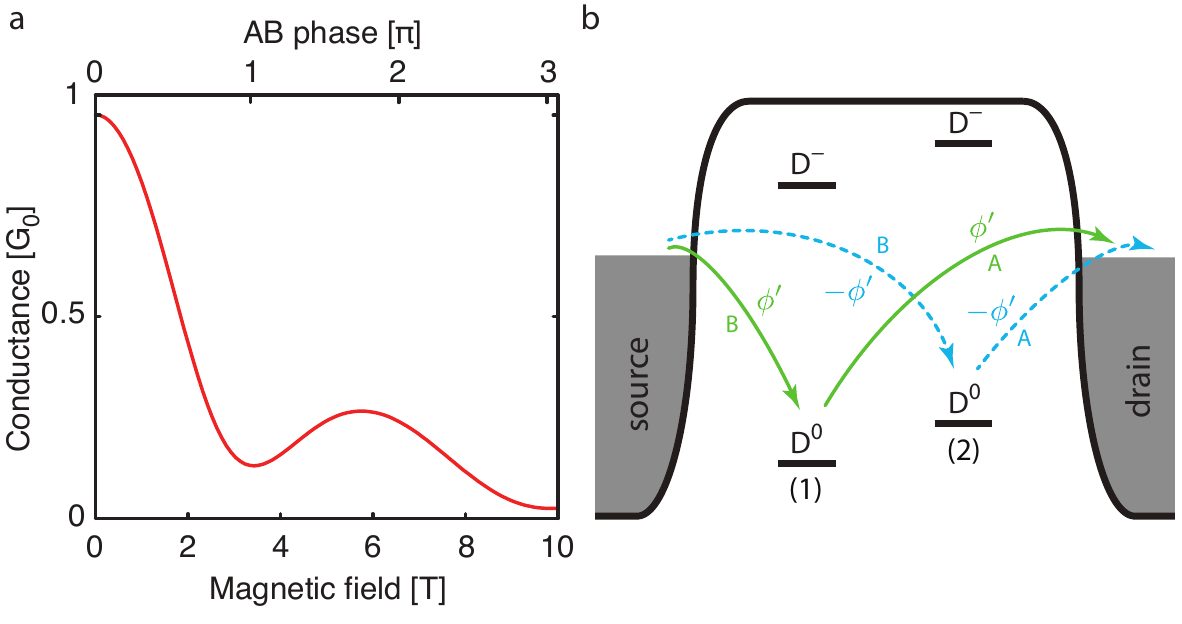}
\caption{ a) The modeled Kondo conductance as a function of magnetic field in the Coulomb blockade region, see Eq.~(\ref{eq:abkondo}) in the main text. On the top axis the corresponding AB phase using a period of 6.5 T is given. The non-monotonic behavior resembles the data in Figure \ref{fig2}a quite closely. b) Two donors are located in the channel of the device and connected to contacts in a parallel arrangement. For simplicity we consider only Kondo tunneling processes via the D$^0$ state. For the example shown, first the localized electron tunnels out to the drain (labeled ``A'') and then an electron tunnels onto the donor from the source (labeled ``B''). The phase due to a finite magnetic field picked up in tunneling, $\phi^\prime$, has the opposite sign for donor (1), in green, and donor (2), in blue.}
\label{fig4}
\end{figure}

To develop a model, we construct a tight-binding Hamiltonian of the two-donor system coupled to contacts, Eq.~(\ref{eq:hamiltonian}). The two arsenic atoms (labeled by $j=1$ and $j=2$) are both represented by single energy level ($\epsilon_j$) quantum dots with an intra-Coulomb interaction ($U_j$), and with $t_j$ the tunneling amplitude to the source ($s$) and drain ($d$) contacts. The contacts are taken as non-interacting reservoirs described by the Hamiltonian $H_{sd}$. The hamiltonian of the complete system is given by:
\begin{eqnarray}
H&=&H_{sd}+\!\!\!\!\!\sum_{j=1,2;\sigma}\!\!\!(\epsilon_j+\sigma\Delta_Z)n_{j,\sigma}+\!\!\!\sum_{j=1,2}\!U_jn_{j\uparrow}n_{j\downarrow} \nonumber \\
&-&\!\!\!\sum_{j=1,2;\sigma}\!\!\!t_{j}\left(e^{i\eta_j\frac{\phi}{4}}|s,\sigma\rangle\langle j,\sigma|+e^{-i\eta_j\frac{\phi}{4}}|d,\sigma\rangle\langle j,\sigma|+h.c.\right).
\label{eq:hamiltonian}
\end{eqnarray}
Here $|s,\sigma\rangle$ and $|d,\sigma\rangle$ are (non-interacting) states in the source and drain contacts respectively (eigenfunctions of $H_{sd}$), $|j,\sigma\rangle$ are the states of the $j^{th}$ donor, ``h.c.'' denotes the hermitian conjugate of the first two terms in the last sum. The number operator, $n_{j,\sigma}$, `measures' the occupation of the $j^{th}$ donor. The effect of the magnetic field is incorporated through the AB phase $\phi=2\pi B/B_{AB}$ picked up during tunneling where $B_{AB}\approx6.5$T and a shift of the resonance away from the Fermi energy by the Zeeman energy $\sigma\Delta_Z=g\mu_{B}B$ ($\sigma=\pm1$) for up spins ($\uparrow$) and down spins ($\downarrow$) \cite{Meir1993,Dong:2001kr}. The parameters $\eta_1=1$ and $\eta_2=-1$ account for the sign difference in the AB phase for different tunneling directions to each of the two donors \cite{Yacoby1995}. We then apply a slave-boson mean field approximation which simplifies our system to a non-interacting Hamiltonian \cite{Dong:2001kr}. The effect of Coulomb interactions is account for by the renormalization of the donors energy levels and their corresponding tunneling rates, denoted with a tilde (e.g. $\tilde{\Gamma_1}$) in the following, as is customary in the framework of slave-boson mean-field approaches \cite{Dong:2001kr}. Using a scattering matrix formalism \cite{Orellana:2006ix} we obtain the following expression for the magnetic field dependent conductance:
\begin{equation}
G(B)=G_{0}\frac{\tilde{\Gamma}_1^2+\tilde{\Gamma}_2^2+2\tilde{\Gamma}_1\tilde{\Gamma}_2\cos(\phi)}{\tilde{\Gamma}_1^2+\tilde{\Gamma}_2^2+\Delta_Z^2+2\tilde{\Gamma}_1\tilde{\Gamma}_2\cos^2(\phi/2)+\tilde{\Gamma}^2_1\tilde{\Gamma}^2_2\sin^4(\phi/2)/\Delta_Z^2}
\label{eq:abkondo}
\end{equation}
with $\tilde{\Gamma}_{1,2}$ the tunnel coupling of the donors to the leads and $G_0$ is the conductance at zero magnetic zero-field. Figure \ref{fig4}a shows a plot of the calculated zero-bias conductance (Eq.~\ref{eq:abkondo}) as a function of magnetic field. In the limit where $\tilde{\Gamma}_{1}=\tilde{\Gamma}_{2}$ and $\Delta_Z\rightarrow \tilde{\epsilon}$, with $\tilde{\epsilon}$ the detuning of the level with respect to the Fermi energy and zero Zeeman shift, this expression is identical to the expression obtained by Lop\'ez {\it et al.} \cite{Lopez:2005cv}. Using parameters similar to the ones in the experiment \cite{KondoGraph}, we obtain a curve that is qualitatively the same as the Kondo data in Figure \ref{fig3}a. Figure \ref{fig4}b shows an example of two second-order tunneling processes that are taken into account in the model. As an electron tunnels on or off a donor, it picks up a phase $\phi^\prime$ (or -$\phi^\prime$) due to the Aharonov-Bohm effect. \par
Comparing the data in Figure \ref{fig3}a and the model in Figure \ref{fig4}a, we notice that at low magnetic field the conductance is less strongly quenched in the experimental data as compared to the theory. This can be understood by the fact that the shift of the Kondo resonances away from zero bias is not exactly twice the Zeeman energy as is assumed for the model. The shift is known to be slightly smaller at fields below the Kondo temperature (i.e. $B<k_{B}T_{K}/g\mu_{B}$) due to correlation effects \cite{Dong:2001kr}. The qualitative agreement with this model, describing the magnetic field dependence, makes us confident that the conductance enhancement at $\sim7.7$ T is indeed the result of the constructive interference of Kondo channels mediated by two donors. Since the enhancement of the Kondo conductance is observed on a similar magnetic field scale as the modulation of the Fano resonance, both effects are most likely caused by the same pair of donors. Furthermore, even though we cannot independently determine whether both donors are actually in the Kondo regime, the agreement between the model and the data indicates that this is indeed the case. \par
\subsection{Phase coherence}
The modulation of the Kondo conductance by the magnetic field through the Aharonov-Bohm effect is a clear experimental proof that the phase is preserved in this system. Our experiment demonstrates that the Aharonov-Bohm can effect persists, even though the phase is carried by a many-body Kondo state in each arm of the system. The role of the Kondo effect is to provide coherent transport channels despite the fact that the sequential transport is completely strongly suppressed due to Coulomb blockade. Since the presence of the AB oscillations are a consequence of the many-body Kondo states being delocalized over both arms into the contacts, there could be a spin-spin interaction present in that region and induce correlations between localized spins \cite{Craig:2004kx,Vavilov:2005bk,Heersche:2006bn}. An interesting open question is whether these kind of interactions are useful in the context of the development of a scalable quantum computer architecture \cite{Bedkihal:2012jg,Tu:2012jm}. \par
Recently demonstrated placement of single donor atoms with atomic precision should allow to fabricate lattices of dopants \cite{Fuechsle:2012hx,Schofield:2003kr}. Using a gate, each donor atom can be tuned such that it holds a single electron at low temperature. Such a structure could, for example, be used to study the formation of exotic collective spin states at low temperature \cite{Aynajian:2012fp,Park:2012go}. This work demonstrates the tunable properties of coupled donors and suggests that they can be used to build a testbed system to study correlation effects. \par
\section{Conclusion}
In conclusion, we present experimental data of an Aharonov-Bohm effect for two donors in a parallel geometry between the source and drain contacts in the regime where transport is mediated by a donor-induced Kondo effect. The Aharonov-Bohm phase is carried by the many-body Kondo states in both arms of the interferometer. This observation is consistent with the phase modulation observed in the sequential transport regime of the same device. A phenomenological model for the Kondo conductance versus magnetic field confirms this by reproducing qualitatively the same trend of the conductance versus magnetic field as the data. These results contribute to the understanding of a system in which the advantageous properties of single donors can be used to create a testbed system to study correlation effects in artificial lattices.
\section*{Acknowledgments}
We are grateful to S. Biesemans and N. Collaert for providing us with the devices. This research was partly conducted by the Australian Research Council Centre of Excellence for Quantum Computation and Communication Technology (project number CE110001027). S.R. acknowledges an ARC Future Fellowship. This work is part of the research program of the Foundation for Fundamental Research on Matter (FOM), which is part of the Netherlands Organization for Scientific Research (NWO).

\section*{References}

\providecommand{\newblock}{}


\begin{thebibliography}{10}
\bibitem{KondoGraph}
{For the plot in Figure 3a the parameters are chosen as $\Gamma_1=0.3$,
  $\Gamma_2=0.12$ and $B_\textrm{AB}=6.5$ T. Where the AB period,
  $B_\textrm{AB}$, was obtained from a fit of the Fano resonance as a function
  of magnetic field.}
\end{thebibliography}
\end{document}